\documentclass[conference]{IEEEtran}
\IEEEoverridecommandlockouts

\usepackage{cite}
\usepackage{amsmath, amssymb, amsfonts}
\usepackage{hyperref}
\usepackage{graphicx}
\usepackage{xcolor}
\usepackage{acro}
\usepackage{cleveref}
\usepackage{verbatim}
\usepackage{subcaption}
\usepackage{float}

\def\BibTeX{{\rm B\kern-.05em{\sc i\kern-.025em b}\kern-.08em
    T\kern-.1667em\lower.7ex\hbox{E}\kern-.125emX}}

\crefformat{equation}{(#2#1#3)}
\crefname{figure}{\figurename}{\figurename}
\crefname{section}{Section}{Sections}
\crefrangeformat{equation}{#3(#1)#4--#5(#2)#6}

\DeclareAcronym{HJB}{
	short=HJB,
	long=Hamilton-Jacobi-Bellman,
}
\DeclareAcronym{LQR}{
	short=LQR,
	long=linear-quadratic regulator,
}
\DeclareAcronym{DDP}{
	short=DDP,
	long=differential dynamic programming,
}
\DeclareAcronym{iLQR}{
	short=iLQR,
	long=iterative linear-quadratic regulator,
}
\DeclareAcronym{SLQ}{
	short=SLQ,
	long=sequential linear-quadratic,
}
\DeclareAcronym{MPC}{
	short=MPC,
	long=model predictive control,
}
\DeclareAcronym{NLP}{
	short=NLP,
	long=nonlinear programming,
}
\DeclareAcronym{PMP}{
	short=PMP,
	long=Pontryagin's minimum principle,
}
\DeclareAcronym{AL}{
	short=AL,
	long=augmented Langrangian,
}
\DeclareAcronym{ODE}{
	short=ODE,
	long=ordinary differential equation,
}
\DeclareAcronym{IVP}{
	short=IVP,
	long=initial value problem,
}
\DeclareAcronym{BVP}{
	short=BVP,
	long=boundary value problem,
}
\DeclareAcronym{EoM}{
	short=EoM,
	long=equations of motion,
}
\DeclareAcronym{FSAL}{
	short=FSAL,
	long=first same as last,
}
\DeclareAcronym{RK}{
	short=RK,
	long=Runge-Kutta,
}
\DeclareAcronym{CDRE}{
	short=CDRE,
	long=continuous-time differential Riccati equation,
}
\DeclareAcronym{DAE}{
	short=DAE,
	long=differential algebraic equation
}

\title{Continuous-time iterative linear-quadratic regulator\\
\thanks{This work was supported by the project 25--16357S of the Czech Science Foundation and by the European Union under the project Robotics and advanced industrial production No. CZ.02.01.01/00/22\_008/0004590. The first and the second author also acknowledge support from the Grant Agency of the Czech Technical University in Prague, student grant No.~SGS23/157/OHK2/3T/12.\\ \\
\textsf{\normalsize Preprint submitted to IEEE}}
}

\author{
\IEEEauthorblockN{
	Juraj Lieskovský,
	Jaroslav Bušek,
	Tomáš Vyhlídal
}
\IEEEauthorblockA{
	Department of Instrumentation and Control Engineering, Faculty of Mechanical Engineering,\\
	Czech Technical University in Prague \\
	}
}

\begin{document}

\maketitle

\begin{abstract}
We present a continuous-time equivalent to the well-known iterative linear-quadratic algorithm including an implementation of a backtracking line-search policy and a novel regularization approach based on the necessary conditions in the Riccati pass of the linear-quadratic regulator. This allows the algorithm to effectively solve trajectory optimization problems with non-convex cost functions, which is demonstrated on the cart-pole swing-up problem. The algorithm's compatibility with state-of-the-art suites of numerical integration solvers allows for the use of high-order adaptive-step methods. Their use results in a variable number of time steps both between passes of the algorithm and across iterations, maintaining a balance between the number of function evaluations and the discretization error.
\end{abstract}


\section{Introduction}

The development of non-linear trajectory optimization methods has been highly impactful in recent years, making approaches such as nonlinear \ac{MPC} of highly non-linear underactuated systems such as legged robots and autonomous drones possible. The problem can be characterized as minimizing the predicted ``cost'' of a system's behavior, when evaluated over a finite time horizon, starting from a single initial condition. Methods for solving this problem can generally be categorized as either direct or indirect. Direct methods transcribe the trajectory optimization problem into a finite-dimensional \ac{NLP} problem, which can then be solved using general-purpose or specialized solvers. Indirect methods, on the other hand, formulate necessary conditions of optimality which a solution to the optimal control problem has to satisfy, dividing each iteration of the algorithm into multiple sub-problems that are solved sequentially.

The most prominent representative of indirect methods is the \ac{DDP} algorithm, derived for systems described in the discrete time domain. Introduced in~\cite{Mayne1966} the algorithm has experienced a resurgence in popularity. Compared to direct methods, it efficiently exploits the problem's Markovian structure and, as a by-product of the optimization process, produces a local feedback controller. The latter allows for the stabilization of precomputed trajectories during execution, which can be taken advantage of in nonlinear \ac{MPC} where the recalculation of the optimal trajectory may be performed over multiple periods of the stabilizing controller's actions.

Due to the time-critical nature of nonlinear \ac{MPC} a variant of the algorithm known as the \ac{iLQR}, first presented in~\cite{Li2004}, has become the standard in solvers~\cite{Howell2019, Mastalli2020, Zhang2025}. Here it is also pertinent to mention the \ac{SLQ} algorithm presented in~\cite{Sideris2005} which is a lesser know equivalent of the \ac{iLQR} algorithm. Compared to the \ac{DDP} algorithm, \ac{iLQR} does not calculate terms containing second-order partial derivatives of the system's dynamics. As a result the algorithm is not guaranteed to converge at the same rate but is computationally faster~\cite{Li2004, Todorov2005}. While originally both the \ac{iLQR} and \ac{SLQ} were derived based on Hamiltonian dynamics, the \ac{iLQR} algorithm was later related to \ac{DDP} and the Bellman principle in~\cite{Tassa2012}.  

A major disadvantage of the \ac{DDP} and \ac{iLQR} algorithms, compared to direct optimization methods, is their inability to handle constraints placed on states and inputs of the system and infeasible initial trajectories. Contemporary research has focused on addressing both. In~\cite{Tassa2014} an approach to handling box-constrained inputs was incorporated, partially addressing the former, while a multiple-shooting variant of the \ac{iLQR} algorithm was introduced in~\cite{Giftthaler2018}, addressing the latter. An \ac{AL} approach to handling equality and inequality constraints was introduced in~\cite{Howell2019}, also allowing for infeasible initial trajectories through the use of slack variables. A similar approach to that in~\cite{Giftthaler2018} was also implemented in~\cite{Mastalli2020} for the full \ac{DDP} algorithm and later extended in~\cite{Mastalli2022} with the input constraint handing introduced in~\cite{Tassa2014}.

The previously mentioned works have focused on optimizing trajectories of systems described in the discrete time domain, resorting to discretizing continuous-time systems using fixed timestep numerical integration. Comparatively, algorithms that produce continuous trajectories and inputs have been overlooked. As one of two notable exceptions, we may consider the \ac{SLQ} algorithm with an \ac{AL} approach to handling equality constraints that was introduced in~\cite{Farshidian2017} and later extended to handle inequality constraints in~\cite{Sleiman2021}. The second exception is the continuous-time \ac{DDP} algorithm presented in~\cite{Sun2014}. Developed to handle terminal constraints in~\cite{Sun2014}, it was later modified for a game-theoretic formulation of the trajectory optimization problem in~\cite{Sun2015}. While these works present interesting problem specific extensions, they do not address crucial topics such as regularization or the problem's numerical properties and efficient implementation.

\subsection{Contributions}


In this paper we provide a concise derivation of a continuous-time equivalent to the standard \ac{iLQR} algorithm, including terms of the full \ac{DDP} algorithm. A novel regularization approach, based on the necessary conditions in \ac{LQR}, is presented and a line-search policy is described. We also discuss details required for efficiently utilizing numerical integration during both the backward and forward pass of the algorithm and address the potential loss of certain properties due to numerical errors. The ability of the algorithm to converge to locally optimal policies is validated on two variants of the cart-pole swing-up problem.

\section{Continuous-time trajectory optimization problem}

The problem of optimizing a continuous-time system's trajectory over a finite horizon can be stated as finding the trajectory $x(\tau) \in \mathbb{R}^{n_x}$ and control policy $u(\tau) \in \mathbb{R}^{n_u}$ that solve the optimization problem
\begin{align*}
\min_{x(\tau), u(\tau)}& \quad \Phi(x(t_f)) + \int_{t_0}^{t_f} l(x(\tau),u(\tau),\tau) \, \mathrm{d}\tau \\
\text{subject to}& \quad \dot{x}(\tau) = f(x(\tau),u(\tau),\tau), \quad \tau \in \langle t_0, t_f \rangle \\
		   & \quad x(t_0) = x_0
\end{align*}
where $x_0$ is the initial state and $f(x,u,t)$ are the continuous-time dynamics of the system. The problem's objective function, made up of the final cost $\Phi(x)$ and the running cost $l(x,u,t)$, is in the context of trajectory optimization referred to as the total cost.

Bellman's principle of optimality introduces additional quantities related to the total cost, namely the cost-go-to and the value function. The cost-to-go is defined as
the ``remaining cost'' from any arbitrary state $x(t)$, $t \in \langle t_0, t_f \rangle$:
\begin{equation}\label{eqn:cost-to-go}
J(x(t),u(\tau),t) = \Phi(x(t_f)) + \int_{t}^{t_f} l(x(\tau),u(\tau),\tau) \, \mathrm{d}\tau
\end{equation}
attained by applying the control policy $u(\tau)$, $\tau \in \langle t, t_f \rangle$. The value function is consequently defined as the cost-to-go for the optimal control policy $u^*(\tau)$ from the state $x(t)$, resulting in the system following the optimal trajectory $x^*(\tau)$:
\begin{equation}\label{eqn:value_function}
V(x(t),t) = J(x^*\!(\tau), u^*\!(\tau), t).
\end{equation}
The value function then figures in the \ac{HJB} equation and the related necessary condition of optimality
\begin{subequations}
\begin{align}
-\frac{\partial V}{\partial t}(x(t),t) &= \min_{u(t)} \, Q(x(t),u(t),t)\label{eqn:HJB} \\
u^*\!(t) &= \arg\min_{u(t)} \, Q(x(t),u(t),t),
\end{align}
\end{subequations}
where
\begin{equation*}
Q(x,u,t) = \\ l(x,u,t) + {\left(\frac{\partial V}{\partial x}\right)}^\top\!\!(x,t) \, f(x,u,t).
\end{equation*}

\section{Proposed Algorithm}\label{sec:algorithm}

Similarly to its discrete-time counterpart, each iteration of our algorithm is centered around a dynamically feasible nominal trajectory $\bar{x}(\tau)$ attained by applying the nominal control policy $\bar{u}(\bar{x}(\tau), \tau)$, where $\tau \in \langle t_0, t_f \rangle$. Every iteration is made up of two passes. A backward pass, during which a quadratic model of the value function is propagated backwards in time while an update to the nominal control policy is created, and a forward pass in which the update is applied to produce a new nominal trajectory.


From this section onward we use the following shorthand: Quantities $f$, $l$, $V$, and $Q$ written with only the argument $t$ represent the given quantity evaluated at $t$, $\bar{x}(t)$, and/or $\bar{u}(t)$. Additionally, the aforementioned quantities written with the argument $t$ and subscripts $t$, $x$, and $u$ denote partial derivatives of the given quantities with respect to those variables evaluated at $t$, $\bar{x}(t)$, and/or $\bar{u}(t)$.

\subsection{Backward Pass}\label{sec:bwd_pass}

Whereas in the discrete-time \ac{DDP} and \ac{iLQR} algorithm the quadratic model of the value function arises from its Taylor expansion, we use the ad-hoc model
\begin{equation}\label{eqn:value_function_model}
V(x,t) = \sigma(t) + s^\top\!(t) \, \delta x(t) + \frac{1}{2} \delta x^\top \! (t) \, S(t) \, \delta x(t),
\end{equation}
where ${\delta x(t)} = x(t) - \bar{x}(t)$ is the difference between the updated and nominal trajectory, $\sigma(t)$ is a scalar function, $s(t)$ a vector valued function, and $S(t)$ a matrix valued function. A local approximation of the problem is then constructed by substituing the model in \cref{eqn:value_function_model} into \cref{eqn:HJB} in addition to the first-order Taylor expansion of $f(x,u,t)$ and second-order expansion of $l(x,u,t)$ around the nominal trajectory:
\begin{align*}
f(x(t),u(t),t) &\approx f(t) + f_x^\top\!(t) \, \delta x(t) + f_u^\top\!(t) \, \delta u(t) \\
l(x(t),u(t),t) &\approx l(t)
+ 
\begin{bmatrix} l_x(t) \\ l_u(t) \end{bmatrix}^\top\! \begin{bmatrix} \delta x(t) \\ \delta u(t) \end{bmatrix} \\ \notag
&\phantom{=}+ 
\frac{1}{2} \begin{bmatrix} \delta x(t) \\ \delta u(t) \end{bmatrix}^\top \begin{bmatrix}  l_{xx}(t) & l_{ux}^\top(t) \\ l_{ux}(t) & l_{uu}(t) \end{bmatrix} \begin{bmatrix} \delta x(t) \\ \delta u(t) \end{bmatrix},
\end{align*}
where ${\delta u(t)} = u(t) - \bar{u}(t)$ is the difference between the updated and nominal control policy. Consequently we attain
\begin{multline}\label{eqn:second_order_HJB}
-\frac{\partial V}{\partial t}(x(t),t) = \\
\min_{\delta u(t)}
\left\{
\frac{1}{2} \begin{bmatrix} 1 \\ \delta x(t) \\ \delta u(t) \end{bmatrix}^\top \! \begin{bmatrix} 2 \, Q(t) & Q_x^\top(t) & Q_u^\top(t) \\ Q_x(t) & Q_{xx}(t) & Q_{ux}^\top(t) \\ Q_u(t)  & Q_{ux}(t) & Q_{uu}(t) \end{bmatrix} \begin{bmatrix} 1 \\ \delta x(t) \\ \delta u(t) \end{bmatrix}
\right\},
\end{multline}
where
\begin{multline*}
\frac{\partial V}{\partial t}(x(t),t) = \dot{\sigma}(t) - s^\top\!(t) \, f(t) + \dot{s}^\top\!(t) \, \delta x(t) \\ - f^\top\!(t) \, S(t) \, \delta x(t) + \frac{1}{2} \delta x^\top\!(t) \, S(t) \, \delta x(t)
\end{multline*}
and
\begin{subequations}\label{eqn:second_order_HJB:terms}
\begin{align}
Q(t) &= l(t) + s^\top\!(t) \, f(t) \\
Q_x(t) &= l_x(t) + S(t) \, f(t) + f_x^\top\!(t) \, s(t) \\
Q_u(t) &= l_u(t) + f_u^\top\!(t) \, s(t) \\
Q_{xx}(t) &= l_{xx}(t) + f_x^\top\!(t) \, S(t) + S(t) \, f_x(t) \label{eqn:Qxx} \\
Q_{ux}(t) &= l_{ux}(t) + f_u^\top\!(t) \, S(t) \\
Q_{uu}(t) &= l_{uu}(t). \label{eqn:Quu}
\end{align}
\end{subequations}
It is important to note that \crefrange{eqn:Qxx}{eqn:Quu} are only approximations of the second-order partial derivatives of $Q(x(t),u(t),t)$, which we define as the objective function of the minimization problem in \cref{eqn:HJB}. Here lies the difference between the proposed \ac{iLQR} algorithm and the full \ac{DDP} algorithm. The \ac{DDP} algorithm uses the full second-order Taylor expansion of $Q(x(t),u(t),t)$, where \crefrange{eqn:Qxx}{eqn:Quu} are replaced by the exact terms 
\begin{align*}
Q^\mathrm{exact}_{xx}(t) &= l_{xx}(t) + f_x^\top\!(t) \, S(t) + S(t) \, f_x(t) + s^\top\!(t) \cdot f_{xx}(t) \\
Q^\mathrm{exact}_{ux}(t) &= l_{ux}(t) + f_u^\top\!(t) \, S(t) + s^\top\!(t) \cdot f_{ux}(t) \\
Q^\mathrm{exact}_{uu}(t) &= l_{uu}(t) + s^\top\!(t) \cdot f_{uu}(t),
\end{align*}
each containing an additional term in the form of a tensor product between $s(t)$ and a second-order partial derivative of the system's dynamics $f_{xx}(t)$, $f_{ux}(t)$, $f_{uu}(t)$, resp. In practice, calculating these terms can be computationally demanding, which is why the \ac{iLQR} algorithm is more commonly implemented in solvers~\cite{Howell2019, Mastalli2020, Zhang2025}.



Assuming $Q_{uu}(t)$ is invertible the minimization problem in \cref{eqn:second_order_HJB} has a unique solution in the form
\begin{equation*}\label{eqn:optimal_policy_update}
\delta u(\delta x(t), t) = -d(t) - K(t) \, \delta x(t),
\end{equation*}
where
\begin{subequations}\label{eqn:optimal_policy_update:terms}
\begin{align}
d(t) &= Q_{uu}^{-1}(t) \, Q_{u}(t) \label{eqn:optimal_policy_update:d} \\
K(t) &= Q_{uu}^{-1}(t) \, Q_{ux}(t) \label{eqn:optimal_policy_update:K}
\end{align}
\end{subequations}
are the feedforward term and the feedback gain. When substituted back into \cref{eqn:second_order_HJB}, we can compare the two sides of the equation in order to attain \aclp{ODE}
\begin{subequations}\label{eqn:value_function_ODE}
\begin{align}
-\dot{\sigma}(t) &= l(t) - \frac{1}{2} Q_u(t) \, Q_{uu}^{-1}(t) \, Q_u(t) \label{eqn:V_ODE_scalar} \\
-\dot{s}(t) &= l_x(t) + f_x^\top\!(t) \, s(t) - Q_{ux}^\top(t) \, Q_{uu}^{-1}(t) \, Q_u(t)\label{eqn:V_ODE_linear} \\
-\dot{S}(t) &= Q_{xx}(t) - Q_{ux}^\top(t) \, Q_{uu}^{-1}(t) \, Q_{ux}(t) \label{eqn:V_ODE_quadratic}.
\end{align}
\end{subequations}
To complete an \ac{IVP} constituting the backward pass we must derive boundary conditions for equations \cref{eqn:value_function_ODE}. From the definition of the value function and cost-to-go in \cref{eqn:value_function} and \cref{eqn:cost-to-go}, we may determine that
\begin{equation}\label{eqn:HJB_boundary}
V(x(t_f),t_f) = \Phi(x(t_f)).
\end{equation}
Consequently, performing a second-order Taylor expansion of \cref{eqn:HJB_boundary}, after substituting in the model from \cref{eqn:value_function_model}, and comparing the two sides of the equation gives us the conditions
\begin{subequations}\label{eqn:value_function_boundary}
\begin{align}
\sigma(t_f) &= \Phi(\bar{x}(t_f)) \\
s(t_f) &= \frac{\partial \Phi}{\partial x}(\bar{x}(t_f)) \\
S(t_f) &= \frac{\partial^2 \Phi}{\partial x^2}(\bar{x}(t_f)). \label{eqn:value_function_boundary:S}
\end{align}
\end{subequations}

The backward pass of the proposed \ac{iLQR} algorithm can be viewed as the Riccati sweep of the finite-horizon continuous-time \ac{LQR}, with an additional bilinear term, for a time-variant system. Therefore, instead of using the standard regularization approach for the \ac{DDP} algorithm as proposed in~\cite{Mayne1966} or the alternative presented in~\cite{Tassa2012}, where a multiple of the identity matrix is iteratively added to the discrete-time equivalent to $Q_{uu}(t)$ or $S(t)$ in both, we propose an approach based on the sufficient conditions for a succesful backward pass of the \ac{LQR}:
\begin{subequations}\label{eqn:reg-cond}
\begin{gather}
l_{uu}(t) \succ 0, \quad \forall t \in \langle t_0, t_f \rangle \label{eqn:reg-cond:luu}\\
\begin{bmatrix} l_{xx}(t) & l_{ux}^\top(t) \\ l_{ux}(t) & l_{uu}(t) \end{bmatrix} \succeq 0, \quad \forall t \in \langle t_0, t_f \rangle \label{eqn:reg-cond:H}\\
\Phi_{xx}(t_f) \succeq 0 \label{eqn:reg-cond:Phi}.
\end{gather}
\end{subequations}
with the consequence of
\begin{equation*}\label{eqn:reg-conseq}
S(t) \succeq 0, \quad \forall t \in \langle t_0, t_f \rangle.
\end{equation*}
As both $l(x(t),u(t),t)$ and $\Phi(x(t),t)$ are not necessarily convex, we may regularize their second-order partial derivatives by minimally modifying their eigen values, a technique standard for quasi-Newton methods in \ac{NLP}~\cite{Nocedal2006}. For a symmetric matrix $A$, its regularization can be achieved in three steps, starting with an eigenvalue decomposition
\[
A = V \Lambda V^\top, \quad \Lambda = \begin{bmatrix} \lambda_1 & & \\ & \lambda_2 & \\ & & \ddots \end{bmatrix}.
\]
Consequently, entries of $\Lambda$ are modified such that
\[
\lambda_{i} := \max(\lambda_{i}, \sqrt{\varepsilon}), \quad \forall i,
\]
where $\varepsilon \in \mathbb{R}$ is a small nonnegative constant and finally $A$ is recomposed. To satisfy both conditions \cref{eqn:reg-cond:luu} and \cref{eqn:reg-cond:H} we may then regularize the matrix in \cref{eqn:reg-cond:H} with $\varepsilon$ set as machine precision.

As is mentioned in~\cite{underactuated}, symmetry of $S(t)$ is sensitive to numerical errors during the integration process of the Riccati sweep. Therefore, to prevent its loss we also utilize the factorized ``square-root'' form
\[
S(t) = P(t) \, P^\top\!(t), \quad \dot{S}(t) = \dot{P}(t) \, P^\top\!(t) + P(t) \, \dot{P}^\top\!(t),
\]
where the factorized version of \cref{eqn:V_ODE_quadratic} takes the form
\begin{equation}\label{eqn:V_ODE_factorized}
-\dot{P}(t) = \frac{1}{2} \left(Q_{xx}(t) - Q_{ux}^\top(t) \, Q_{uu}^{-1}(t) \, Q_{ux}(t) \right) {P(t)}^{-\!\top}.
\end{equation}
Since $P(t)$ must be invertible $\forall t \in \langle t_0, t_f \rangle$, $\Phi_{xx}(t_f)$ must satisfy a more strict condition than stated in \cref{eqn:reg-cond:Phi}, of being positive-definite. This can be ensured using a similar regularization process as described earlier, during the formation of the boundary condition
\begin{equation}\label{eqn:factor:Pf}
P(t_f) = V \Lambda^{\frac{1}{2}}, \quad \Lambda^{\frac{1}{2}} = \begin{bmatrix} \sqrt{\lambda_1} & & \\ & \sqrt{\lambda_2} & \\ & & \ddots \end{bmatrix},
\end{equation}
by modifying the entries of $\Lambda^{\frac{1}{2}}$ such that
\[
\sqrt{\lambda_{i}} := \max(\sqrt{\lambda_{i}}, \sqrt[4]{\varepsilon}), \quad \forall i,
\]
with $\varepsilon$ set as machine precision.

\subsection{Forward Pass}\label{sec:fwd_pass}

In the forward pass, the dynamics of the system are integrated from the initial conditions $x(t_0) = x_0$ while applying the updated control policy. As we have discussed in the previous section, $\delta u(\delta x(t),t)$ corresponds to the full step of a quasi-Newton method. For this reason we apply a backtracking line-search procedure in the forward pass, forming the updated control policy as
\begin{equation}
u(x(t),t,\alpha) = \bar{u}(t) - \alpha \, d(t) - K(t) (x(t) - \bar{x}(t)), \label{eqn:policy}
\end{equation}
where $\alpha \in (0, 1\rangle$ is the line-search parameter. Associated with the updated control policy is the expected improvement, i.e.\ the difference between the quadratic model of the value function at the start of the trajectory and the cost of the nominal trajectory
\begin{equation*}
\Delta V(\alpha) = -\left(\alpha - \frac{\alpha^2}{2}\right) \int_{t_f}^{t_0} Q_u^\top\!(t) \, Q_{uu}^{-1}(t) \, Q_{u}(t) \, \mathrm{d}t .
\end{equation*}

At the start of the line-search procedure $\alpha$ is initialized at the value $\alpha = 1$ and a rollout of the system's dynamics with the updated policy in \cref{eqn:policy} is performed. If the condition 
\begin{equation}\label{eqn:success_cond}
J(x(t),u(x(t),t,\alpha)) - J(\bar{x}(t),\bar{u}(t)) < \beta \, \Delta V(\alpha),
\end{equation}
where $\beta \in \mathbb{R}$ is a small positive constant, e.g.\ $10^{-4}$, similar to the Armijo condition is satisfied the pass is determined as successful and the updated trajectory and policy are accepted as the new nominal trajectory and policy for the following iteration. If \cref{eqn:success_cond} is not satisfied, $\alpha$ is reduced by a factor $\rho \in (0, 1\rangle$ and the rollout is repeated. This process is repeated until a pass is successful or $\alpha$ decreases below a certain threshold $\alpha_{\min}$, at which point the algorithm is terminated.

\section{Numerical Integration}

As the \ac{IVP} value problems described in \cref{sec:algorithm} do not have an analytical solution, numerical integration must be used. The problem in the algorithm's backward pass is not only stiff, but its stiffness also varries greatly across the horizon, as will be shown in \cref{sec:results}. Consequently, the application of the feedback policy that is produced then results in a similar behavior during the forward pass. Therefore an integration method for stiff problems with adaptive time-stepping is appropriate. Independently, non-state variables $d(t)$ and $K(t)$ from the backward pass are re-used in the forward pass and $u(t)$, calculated during the forward pass, is in-turn re-used during the backward pass. Therefore, it is beneficial to simulate both the forward and backward pass as a \ac{DAE} problem with the aformentioned quantities included as algebraic variables and use an integration method that provides dense output for their interpolation. Suitable from both aspects are Rosenbrock-Wanner methods with embedded lower-order methods for step-size adaptation and dense output formulas, such as the Rodas5P method presented in~\cite{Steinebach2023}.

As for the particular \acp{DAE} forming the problems of the algorithm's backward and forward pass, in the backward pass, the algebraic equations are \crefrange{eqn:optimal_policy_update:d}{eqn:optimal_policy_update:K} and for the differential equations we manipulate \crefrange{eqn:V_ODE_scalar}{eqn:V_ODE_linear} and \cref{eqn:V_ODE_factorized} into the form
\begin{align*}
-\dot{\sigma}(t) &= l(t) - \frac{1}{2} d^\top\!(t) \, Q_{u}(t) \\
-\dot{s}(t) &= l_x(t) + f_x^\top\!(t) \, s(t) - K^\top\!(t) \, Q_u(t) \\
-\dot{P}(t) &= \frac{1}{2} \left(Q_{xx}(t) - K^\top\!(t) \, Q_{uu}(t) \, K(t) \right) {P(t)}^{-\!\top} ,
\end{align*}
with boundary conditions from \cref{eqn:value_function_boundary} where \cref{eqn:value_function_boundary:S} is factorized according to \cref{eqn:factor:Pf}. In the forward pass the problem is made up of the algebraic equation \cref{eqn:policy} and differential equations
\begin{align*}
\dot{x}(t) &= f(x(t),u(t),t) \\
\dot{c}(t) &= l(x(t),u(t),t),
\end{align*}
with initial conditions $x(t_0) = x_0$ and $c(t_0) = 0$, where 
\begin{equation*}
c(t) = \begin{cases}
	\int_{t_0}^t l(x(\tau),u(\tau),\tau) \, \mathrm{d}\tau,& \quad t \in \langle t_0, t_f) \\
	J(x(\tau), u(\tau)),& \quad t = t_f
\end{cases}
\end{equation*}
is the accumulated cost.

\section{Numerical Validation}\label{sec:results}

To validate the proposed algorithm, two variants of the cartpole swing-up problem were solved. The algorithm's implementation was written in the julia programming language using the \verb|DifferentialEquations.jl|~\cite{Rackauckas2017} suite of solvers for numerical integration. From the suite we selected the fifth-order Rosenbrock-Wanner method, presented in~\cite{Steinebach2023}, to solve both passes of the algorithm. For the two variants of the problem, the algorithm's parameters were set to $\varepsilon = 2^{-52}$ (64bit machine precision), $\beta = 10^{-4}$, $\rho = \frac{1}{2}$, and $\alpha_{\min}=2^{-20}$.

\begin{figure}[tb]
	\centering
	\includegraphics[width=60mm]{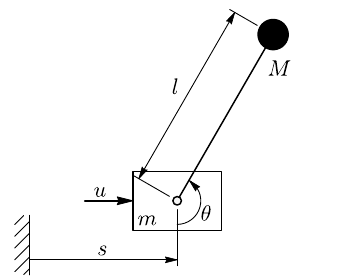}
	\caption{Diagram of the cart-pole system}\label{fig:cartpole:diagram}
\end{figure}

A simple model of the cart-pole system, in which a point mass is placed at the end of a pole, whose inertial properties are neglected, was used. Parameters of the model, depicted in \cref{fig:cartpole:diagram}, were chosen as $m = 1$, $l=0.2$, $M = 0.1$, and $g=9.81$. The states of the system were chosen as the position of the cart $s(t)$, the angle of the pole $\theta(t)$, and their derivatives with respect to time $\dot{s}(t)$ and $\dot{\theta}(t)$:
\[
x(t) = \begin{bmatrix} s(t) & \theta(t) & \dot{s}(t) & \dot{\theta}(t) \end{bmatrix}^\top.
\]
A force acting in the direction of the coordinate $s(t)$ was then chosen as the only input of the system. Both variants of the problem were solved with the horizon of $t \in \langle 0, 2\rangle$, the initial state
\[
x_0 = \begin{bmatrix} 0 & 10^{-3} \pi & 0 & 0 \end{bmatrix}^\top\!,
\]
to avoid the singularity at $x(t_0) = 0$, and the nominal control policy $\bar{u}(t) = 0$, $\forall t$. The two solved variants of the problem then differ only in the definition of the running and final cost, the first using a convex running and final cost, and the second only a non-convex running cost.

\subsection{Convex Problem}

\begin{figure}[tb]
	\centering
	\begin{subfigure}[b]{\linewidth}
	\includegraphics[width=\textwidth]{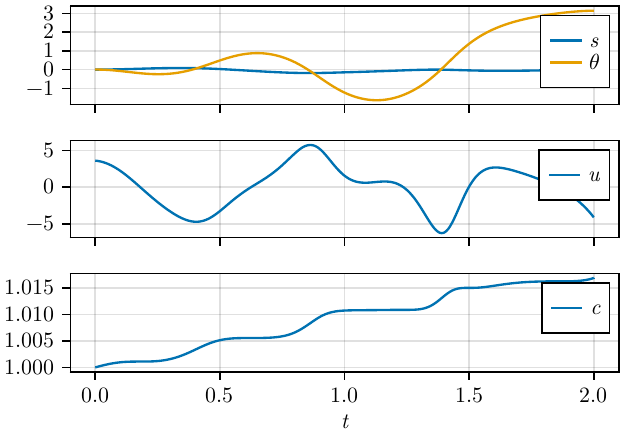}
	\caption{Resulting trajectory of states $s$ and $\theta$, control policy $u$, and accumulated cost $c$}\label{fig:cartpole:convex:trajectory}
	\vspace*{1em}
	\end{subfigure}
	\begin{subfigure}[b]{\linewidth}
	\includegraphics[width=\textwidth]{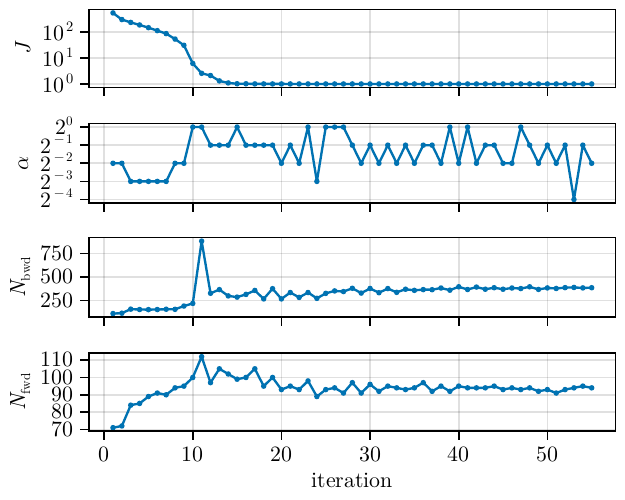}
	\caption{Total cost $J$, line-search parameter $\alpha$, and number of steps taken in the backward pass $N_{\text{bwd}}$ and forward pass $N_{\text{fwd}}$ of the algorithm in each iteration.}\label{fig:cartpole:convex:iterations}
	\end{subfigure}
	\caption{Results for the variant of the problem with convex cost functions.}
\end{figure}

The first variant of the problem corresponds a fairly standard time-constrained minimal-energy swing-up with a quadratic running cost
\[
l(x(t),u(t),t) = 10^{-3} u^2(t)
\]
and a final cost in the form of a quadratic penalty function around the desired final state:
\begin{multline*}
\Phi(x(t)) = 10^{2} \left(s^2(t) + {(\theta(t)-\pi)}^2\right) + \dot{s}^2(t) + \dot{\theta}^2(t).
\end{multline*}
The algorithm converged to the locally optimal trajectory with a total cost of 1.016 with timesteps in the range of $\langle 10^{-6}, 1.5 \cdot 10^{-2} \rangle$ during the backward pass and $\langle 10^{-6}, 6.9 \cdot 10^{-2} \rangle$ during the forward pass. The resulting trajectory is shown in \cref{fig:cartpole:convex:trajectory} along with the control policy and accumulated cost. Convergence was achieved in 55 iterations. The total cost, line-search parameter, and number of steps taken during the backward and forward pass of the algorithm are plotted separately for each iteration in \cref{fig:cartpole:convex:iterations}. In \cref{fig:cartpole:convex:trajectory}

\subsection{Non-Convex Problem}

To demonstate the algorithms ability to solve trajectory optimization problems with non-convex cost functions, the second variant of the problem made use of a manually tuned \acs{MPC}-like running cost
\begin{multline*}
l(x(t),u(t),t) = 1 + \cos(\theta(t)) + 10 \, s^2(t) + 3 \cdot 10^{-2} u^2(t),
\end{multline*}
where $1 + \cos(\theta(t))$ is proportionate to the potential energy of the system, and no final cost.
 
\begin{figure}[tb]
	\begin{subfigure}[b]{\linewidth}
	\includegraphics[width=\textwidth]{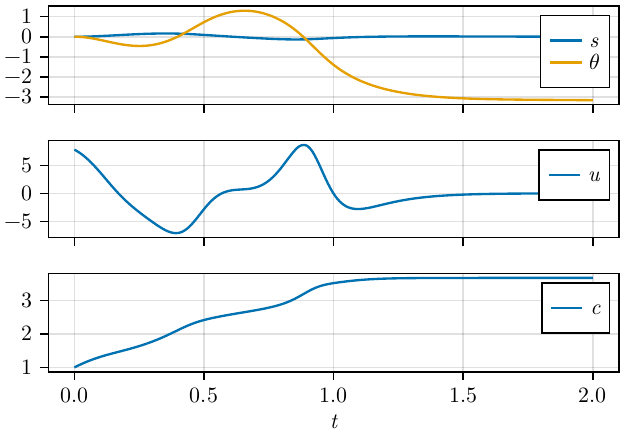}
	\caption{Resulting trajectory of states $s$ and $\theta$, control policy $u$, and accumulated cost $c$}\label{fig:cartpole:nonconvex:trajectory}
	\vspace*{1em}
	\end{subfigure}
	\begin{subfigure}[b]{\linewidth}
	\includegraphics[width=\textwidth]{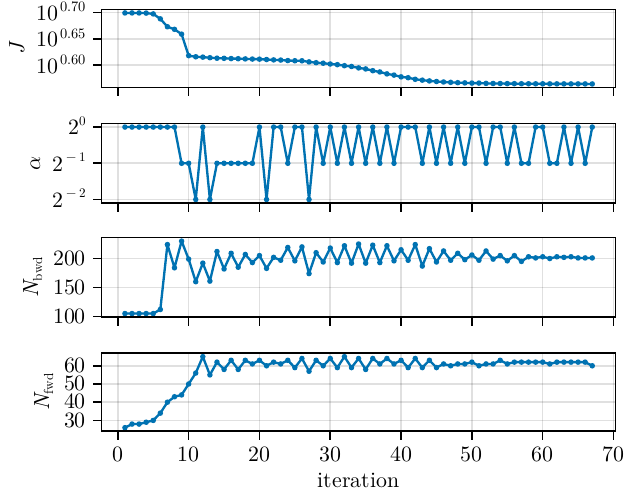}
	\caption{Total cost $J$, line-search parameter $\alpha$, and number of steps taken in the backward pass $N_{\text{bwd}}$ and forward pass $N_{\text{fwd}}$ of the algorithm in each iteration.}\label{fig:cartpole:nonconvex:iterations}
	\end{subfigure}
	\caption{Results for the variant of the problem with a non-convex running cost.}
\end{figure}

The algorithm converged to the locally optimal trajectory with a total cost of 3.668 with timesteps in the range of $\langle 3.1 \cdot 10^{-9}, 3.5 \cdot 10^{-2} \rangle$ during the backward pass and $\langle 10^{-6}, 9.7 \cdot 10^{-2} \rangle$ during the forward pass. The resulting trajectory is shown in \figurename~\ref{fig:cartpole:nonconvex:trajectory}, along with the control policy and the accumulated cost. Convergence was achieved in 67 iterations. The total cost, line-search parameter, and number of steps taken during the backard and forward pass of the algorithm are plotted separately for each iteration in \cref{fig:cartpole:nonconvex:iterations}.

\section{Conclusion}

We have presented a continuous-time variant of the \ac{iLQR} algorithm, including a novel regularization scheme and a line-search policy. Together they ensure the algorithm converges to a local minimum not only for convex but also nonconvex cost functions, which was validated in simulation. The algorithm's backward and forward pass are formulated as \ac{DAE} \acp{IVP}, allowing for the use of adaptive-step integration methods with dense output. The utility of their use was demonstated during the numerical validation, where the \acp{IVP} exhibited highly variable stiffness.

Similarities with the (discrete-time) \ac{iLQR} algorithm offer the posibility of implementing extensions for incorporating state and input constraints, infeasible initial trajectories, etc. In combination with the unique property of handling (variably) stiff dynamics in both the forward and backward pass, implementation of these extensions could further enlarge the set of problems solvable using trajectory optimization.

\bibliographystyle{IEEEtran}
\bibliography{IEEEabrv,references}

\end{document}